\documentclass[a4paper]{article}
\usepackage{graphicx}
\usepackage{natbib}
\usepackage{graphics}
\newcommand{\tablenotemark}[1]{${}^{#1}$}
\newcommand{\tablecomments}[1]{}
\newcommand{\tablenotetext}[2]{${}^{#1}${#2}}

\newcommand{\aj}{AJ}
\newcommand{\araa}{ARA\&A}
\newcommand{\apj}{ApJ}
\newcommand{\aap}{A\&A}
\begin{document}


\title{Observations of type 1a supernovae are consistent with a static universe}
\author{David F. Crawford\\
Sydney Institute for Astronomy, School of Physics, University of Sydney\\
Address: 44 Market St, Naremburn, 2065, NSW, Australia.\\ davdcraw@bigpond.net.au}
\maketitle

\begin{abstract}
Analysis of type 1a supernovae observations out to a redshift of $z$=1.6 shows that there is  good agreement between the light-curve widths and $(1+z)$ which is usually interpreted as a strong support for time dilation due to an expanding universe. This paper argues that a strong case can be made for a static universe where the supernovae light-curve-width dependence on redshift is due to selection effects. The analysis is based on the principle that it is the total energy (the fluence) and not the peak magnitude that is the best `standard candle' for type 1a supernovae. A simple model using a static cosmology provides an excellent prediction for the dependence of light curve width on redshift  and the luminosity-width relationship for nearby supernovae. The width dependence arises from the assumption of constant absolute magnitude resulting in strong selection of lower luminosity supernovae at higher redshifts due to the use of an incorrect distance modulus. Using a static cosmology, curvature-cosmology, and without fitting any parameters the analysis shows that the total energy is independent of redshift and provides a Hubble constant of $63.1\pm2.5$ kms$^{-1}$ Mpc$^{-1}$. There is no indication of any deviation at large redshifts that has been ascribed to the occurrence of dark energy.
keywords{cosmology: observations, large-scale structure of universe, theory,  supernovae}
\end{abstract}

\section{Introduction}
The type 1a supernovae are believed to be the result of an explosion of a white dwarf that has been steadily acquiring matter from a close companion. When the mass exceeds the Chandrasekhar limit, the white dwarf explodes producing a very bright supernova whose light curve shows a rapid rise over several weeks, then an equally rapid fall followed by a much slower decrease over several hundred days. The type 1a supernovae are distinguished from other types of supernovae by the absence of hydrogen lines and the occurrence of strong silicon lines in their spectra near the time of maximum luminosity. Although the theoretical modelling is poor, there is much empirical evidence, from nearby supernovae, that they all have remarkably similar light curves, both in absolute magnitude and in their time scales. This has led to a considerable effort to use them as cosmological probes. Since they have been observed out to redshifts with $z$ greater than one they have been used to test the cosmological time dilation that is predicted by expanding cosmologies. Support for the lack of time dilation in a static universe also comes from a study of Gamma Ray Bursts \citep{Crawford09a,Crawford09b}.

Several major projects have used both the Hubble space telescope and large earth-bound telescopes to obtain a large number of type 1a supernova observations, especially to large redshifts. They include the Supernova Cosmology Project \citep{Perlmutter99,Goldhaber01,Knop03}, the Supernova Legacy Survey \citep{Astier05}, the Hubble Higher z supernovae Search \citep{Strolger04} and the ESSENCE supernova survey \citep{Wood-Vasey07, Davis07}. Recently \citet{Kowalski08} have provided a re-analysis of these survey data and  all other relevant supernovae in the literature and have provided new observations of some nearby supernovae. Because these {\em Union} data are comprehensive, uniformly analysed and include nearly all previous observations, the following analysis will be confined to these data. The selected data provide (Bessel) B-band magnitudes and stretch factors, and $B-V$ colours for supernovae in the range $0.015\le z\le 1.551$.

\begin{figure}[!htb]
\includegraphics[width=0.99\textwidth,clip=true]{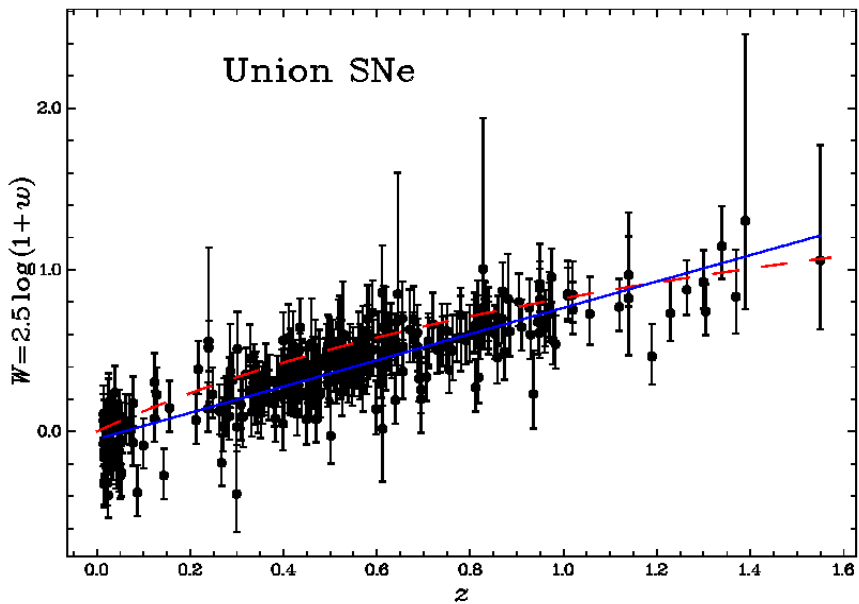}
\caption{Width of supernovae type 1a verses redshift for Union data. The solid line (blue) is expected time dilation in an expanding cosmology.  The dashed line (red) is the function $0.79f(z)$ (Equation \ref{dm}). The error bars show the final corrected uncertainties. \label{f6}}
\end{figure}

The widths (relative to the standard width) of the supernova light curves are derived from the stretch factors $s$, provided by \citet{Kowalski08} by the equation
\begin{equation}
\label{e109}
w = (1+z)s.
\end{equation}
It should be noted that the widths are the measured value and the stretch values were derived from the widths. The uncertainty in each width was taken to be $(1+z)$ times the quoted uncertainty in the stretch value. Figure \ref{f6} shows a plot of these widths as a function of redshift. These results, which show excellent quantitative agreement with the predicted time dilation, have been hailed as one of the strongest pieces of evidence for an expanding cosmology.

\section{Uncertainties and rejections}
Table \ref{sna1} shows as a function of redshift the actual root-mean-square (rms)  and the average quoted uncertainty for the stretch and magnitude variables. The rms values are derived from the residuals after removing a fitted power law as a function of $(1+z)$. This was done to remove any simple cosmological dependencies. In order to see whether the uncertainties showed any redshift dependence the original 306 top quality supernovae \citep{Kowalski08} were divided into 6 cuts with approximately the same number of supernovae in each cut. The cut boundaries were at $z$=0.015, 0.15, 0.4, 0.5, 0.62, 0.82 and 1.6. If the quoted uncertainties are realistic  estimates then there should be reasonable agreement between them and the observed rms values. That is the numbers in column three should roughly agree with those in column 4 and similarly columns five and six should be in agreement. Clearly there is only a small redshift dependence but a large difference between the measurement uncertainties and the observed rms values.
\begin{table}[!htb]
\begin{center}
{\caption{Uncertainty analysis}\label{sna1}}
\begin{tabular}{lccccccc}
\hline
No. & $z$ & $s_{rms}$ \tablenotemark{a} & $\overline{\sigma_s}$ \tablenotemark{b} & $M_{rms}$ \tablenotemark{c} & $\overline{\sigma_M}$ \tablenotemark{d} & \\
54  & 0.034 & 0.156 & 0.028 & 0.244 & 0.087 \\
50  & 0.317 & 0.173 & 0.053 & 0.276 & 0.073 \\
50  & 0.449 & 0.123 & 0.062 & 0.275 & 0.068 \\
51  & 0.561 & 0.157 & 0.076 & 0.258 & 0.080 \\
45  & 0.713 & 0.143 & 0.073 & 0.295 & 0.091 \\
50  & 1.004 & 0.156 & 0.110 & 0.271 & 0.133 \\
All & 0.505 & 0.153 & 0.066 & 0.274 & 0.089 \\
\hline
\end{tabular}
\end{center}
\tablenotetext{a}{The root-mean-square of residuals for stretch (s)}\\
\tablenotetext{b}{The average quoted uncertainty for stretch (s)}\\
\tablenotetext{c}{The root-mean-square of residuals for magnitude (M)}\\
\tablenotetext{d}{The average quoted uncertainty for magnitude (M)}\\
\end{table}

Assuming that the differences are due to an unknown additional variation the quoted uncertainties have been increased by adding (in quadrature) 0.139  to the stretch uncertainties and 0.118 to the  magnitude uncertainties. These values gave good agreement between the rms values and the new uncertainties and the combined uncertainties are used in this analysis. In effect the uncertainties are used only to provide relative weights. The final uncertainties in measure parameters are scaled so that the residual $\chi^2$ value is equal to the number of degrees of freedom. Thus the uncertainties in the computed parameters include contributions from the measurement uncertainties, the intrinsic scatter and the inaccuracies in the fitted model.

The next step was to examine individual supernovae and to reject any that had, after removing any redshift dependence, a residual $\chi^2_1$ that was greater than 10.83 which corresponds to a probability of 0.001. The supernovae 1997o, 03D1fq and 2002W were rejected because of anomalous stretch values and 1998co, 2001jf, and 1999gd were rejected because of anomalous magnitudes.

\section{The supernova model}
Since the time dilation is intrinsic to all expansion cosmologies it is essential that a test of time dilation must not depend either directly or indirectly on an expansion cosmology. For type 1a supernovae it will be shown that the choice of cosmology does have an important effect on the selection of supernovae and the use of concordance cosmology leads to a biased sample. As shown in Figure \ref{f6} the fit of the light-curve widths for this sample is in good agreement with time dilation. It will be argued that this agreement arises from the way in which the supernovae are selected. This will be done using a static cosmology, curvature-cosmology (briefly described in the appendix).

The current model for type 1a supernovae \citep{Hillebrandt00} is that a white dwarf in a binary system gradually accretes matter from its companion. Eventually it exceeds the limits of its stability - the Chandrasekhar limit - and there is a thermonuclear incineration of the white dwarf. Because this is a well-defined limit to the supernova mass and hence its energy, it is expected that each supernova has, on average, the same total energy output. The light curve has a rise time of about 20 days followed by a fall of about 20 days and then a long tail that is most likely due to the decay of $^{56}$Ni. The widths are measured in the light coming from the expanding shells before the radioactive decay dominates. Thus the widths are a function of the structure and opacity of the initial explosion and have little dependence on the radioactive decay.

The proposed model is based on the principle that the most unchanging characteristic of type 1a supernovae is their total energy and not their peak magnitude. Due to local effects the total energy will have small variations about a constant value.  Since the total energy of a supernova is proportional to the area under its light curve it is proportional to the product of the maximum luminosity and the width of the light curve. Thus constant total energy means that a selection of supernovae based on magnitude will automatically produce a selection in width.

Since the distance modulus for  concordance  cosmology is always larger than that for curvature-cosmology as the redshift increases a selection based on absolute magnitude that uses concordance cosmology will select fainter supernovae and hence supernovae with wider light curves. The effect of this selection will increase with redshift. It will be shown that this model can explain the results shown in Figure \ref{f6}. Moreover it can explain more detailed results about the magnitude-width relationship and make reasonable predictions for the observed number of supernovae that have been selected as a function of redshift. Finally it is shown that the average total energy is independent of redshift.

\section{Analysis for constant energy}
\label{s6.5.1}
For consistency with the magnitudes and to make it easier to interpret straight lines as power laws, define a new variable $W$ by $W = 2.5\log(w)$. Except for a multiplying constant (with a value of 1.086) $W$ is essentially identical to $w$ for small widths. Similarly a new redshift variable, $Z$, is defined by $Z=2.5\log(1+z)$.

Since the redshift in curvature-cosmology arises from an interaction with the intervening gas, it is not always a good measurement of distance. In particular the halo around our galaxy and that around any target galaxy will produce an extra redshift that results in an overall redshift that is larger than would be expected from the distance and a constant inter-galactic gas density. Since this is an additive effect it is important only for nearby objects. In fact the nearby supernovae (defined as those with $z<0.15$) show an average absolute magnitude that is brighter than the extrapolated magnitude from more distant galaxies. In order to make a partial correction for this bias all redshifts were reduced by subtracting 0.006 from each redshift, $z$. This correction brought the magnitudes into agreement.

The supernova model interprets the apparent increase of light-curve width with redshift as shown in Figure \ref{f6} as being due to selection of the supernovae. This interpretation depends on the results that follow. First, we examine the dependence of absolute magnitude on width and find a strong relationship that is opposite to that which is currently accepted \citep{Phillips93}, \citet{Hamuy96}. On the other hand, it is consistent with a physical model in which the type 1a supernovae can have variable widths and absolute magnitudes subject to the constraint that the total energy output is approximately constant. Second, since the supernovae are, in part, selected by the requirement that their absolute magnitude computed using the concordance distance modulus is close to a reference value, then as the redshift increases intrinsically fainter supernovae are selected. If constant energy applies the fainter magnitudes will be associated with greater widths in a manner that mimics time dilation.

For curvature-cosmology, the distance modulus is given by Equation (\ref{e8}) that has only one free parameter, the Hubble constant and since this is an additive constant, its actual value is not important in this context. However, in order to have realistic magnitudes for the plots, the analysis is done with $h=0.7$ (i.e. $H=70$ km s$^{-1}$Mpc$^{-1}$). Since there is a very small but significant dependence of the absolute magnitude on the $B-V$ colour following \citet{Kowalski08} the magnitudes were reduced by a term $\beta(B-V)$ where $\beta$ was determined by minimising the $\chi^2$ of the residuals after fitting $M$ verses $W$. This gave $\beta=1.15$ which can be compared with $\beta=2.28$ given by \citet{Kowalski08}.

\begin{figure}[!htb]
\includegraphics[width=0.99\textwidth,clip=true]{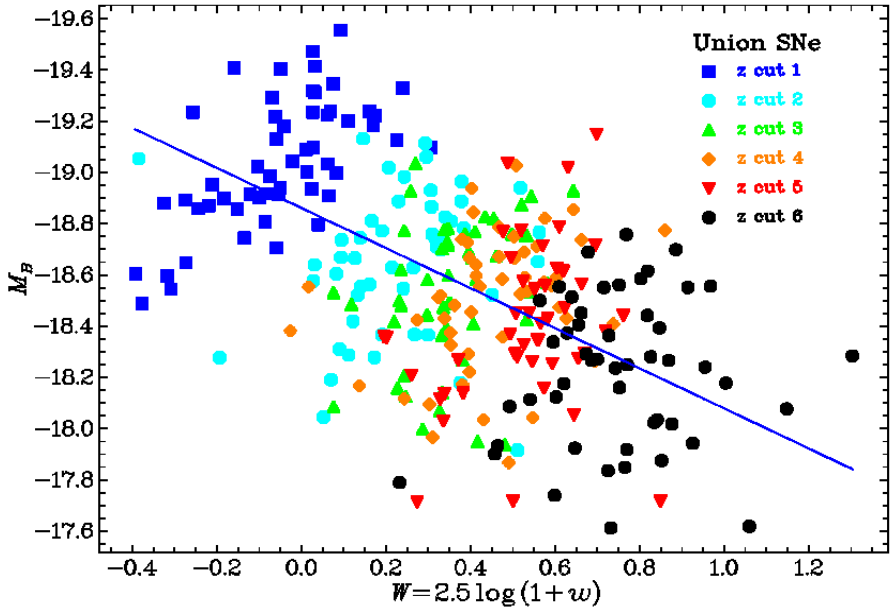}
\caption{Absolute magnitudes (Bessel B-band) of type 1a supernovae verses widths of their light curves in curvature-cosmology for Union data. The solid line (blue) is the line of best fit. The data are divided into 6 redshift cuts corresponding to the cuts shown in Table \ref{sna1}. Note that the brightest supernovae are at the top.\label{f7}}
\end{figure}

A plot of absolute magnitude (using curvature cosmology) verses width is shown in Figure \ref{f7}. For later analysis the data are divided into 6 redshift cuts corresponding to the cuts shown in Table \ref{sna1}. The best-fit (using the subroutine {\sc fitexy} provided by \citet{Press92}) straight line has a slope of $0.877\pm0.085$.  Thus supernovae that are brighter have narrower widths, or the weaker are wider.

To assess whether this dependence of absolute magnitude on width shown in Figure \ref{f7} is realistic it is necessary to understand what it physically means. The fitted slope shows that the luminosity - width relationship is
\begin{equation}
\label{e111}
L \propto w^{-0.877\pm0.085}.
\end{equation}
which since the total energy is $E \propto Lw \propto w^{0.123\pm 0.085}$, it agrees with the model of a constant total energy. We may reasonably conclude that this relationship is consistent with the known physics of the explosion. The total energy emitted is often called the (energy) fluence. Thus define the fluence by $S=-2.5\log(E/E_0)$ where $E$ is the total energy and $E_0$ is the total energy emitted by a reference supernova with unit width at a distance of 10 pc. In general there will be colour-band or bolometric corrections. But since only the Bessel B-band is considered here these corrections are not needed.  Thus the fluence for a supernova is given by $S=M-W$. Since, by definition, the reference width, $W$, is zero the relationship between the absolute magnitude and the width can be expressed as $W=M-M_{ref}$, where $M_{ref}$ is the absolute magnitude for the standard fluence.

The result that the power law shown by Equation (\ref{e111}) has a slope consistent with one means that the property that is common to all type 1a supernovae is their total energy of emission, the fluence.  Therefore, we can use the fluence as a `standard candle'. These results are essentially identical in form to those for GRB which were shown \citep{Crawford09b} to be more consistent with constant energy rather than constant luminosity.

\section{Correlations between widths and magnitudes}
It is well known \citep{Phillips93} that there is a correlation between the width of the supernovae light curve and its absolute magnitude. However, the dependence found was obtained from the local supernovae and has the opposite sense to the slope shown by the solid line in Figure \ref{f7}.  If we group the supernovae into redshift bands as shown by the difference symbols (and colours) in Figure \ref{f7} it is clear that within each band the magnitude verses width dependence has a slope of the opposite sign. In fact the first cut for nearby supernovae with $0.015\le z \le 0.15$ (the blue squares) has a slope of $-1.33\pm0.21$. This is in good agreement with $-1.56\pm0.25$ reported by \citet{Guy05} using similar recent data.

If this opposite dependence is not explained it would be a strong argument against the constant energy hypothesis. Since the supernovae will differ in composition and external influences, such as external pressure, they are not expected to have exactly the same fluence. The extension to the proposed model is that the fluence has a variation that effects both the magnitude and width but that the shape of the light curve remains unaltered as the fluence varies. For example suppose the fluence is greater than the standard value by $\Delta$. Then to keep the shape of the light curve proportional to the reference light curve the change is spread equally between the luminosity and the width. Thus the magnitude, $M$, decreases by $\Delta/2$ and the width, $W$, increases by $\Delta/2$. Thus for a fixed redshift the magnitude will be a function of the width, $W$, with a negative sign. This qualitatively explains the dependence reported by  \citep{Phillips93}.  An unbiased selection of supernovae would have the same expected fluence at all redshifts  and the global slope would be negative. However the selection bias has produced a sample in which the variation in magnitude due to redshift is larger than the intrinsic variation in magnitude due to the variation in fluence. This produces a variation that overwhelms the intrinsic variation to produce a positive global slope. In effect there is an intrinsic negative slope that is produced by the variations in fluence with a constant shape for the light curve and because of biased selection there is a positive slope of the distribution of average magnitude at each redshift as a function of redshift.

\begin{figure}[!htb]
\includegraphics[width=0.99\textwidth,clip=true]{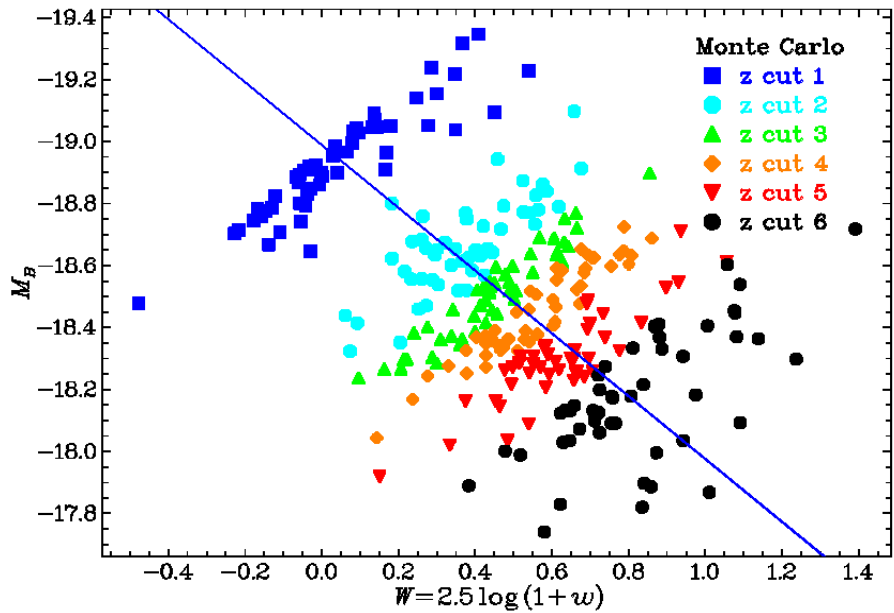}
\caption{Absolute magnitude verses width for a Monte Carlo simulation. Solid line (blue) is the line of best fit.\label{f8}}
\end{figure}

In order to demonstrate this effect and to illustrate the overall selection process, a Monte Carlo simulation was used that preserved the individual redshifts but had completely random widths and magnitudes. Keeping the redshifts the same ensues that the simulation has the identical redshift distribution. Let $\xi$ be a random variate drawn from a normal distribution with mean of zero and standard deviation equal to $\sigma$. Then for each redshift a random variate, $\xi$, was drawn with standard deviation 0.16 mag. Using this variate the width and magnitude for a supernova at that redshift were computed by $W=f(z)-\xi$ and $M=M_{ref}+f(z)+\xi$ where $f(z)$ is defined by Equation (\ref{dm}). This was repeated for each supernova redshift to provide a full set that was analysed identically to the observed data. One thousand sets were used to increase the precision in computed parameters. Note that the random variate $\xi$ that is used for $W$ is used with the opposite sign for $M$. This provides a linear dependence with a slope of minus one. In contrast $f(z)$ is added with the same sign. This produces a global linear dependence with a slope of plus one. Figure \ref{f8} shows a typical set. It has the same basic structure as the observed data shown in Figure \ref{f7}.

As noted the observed slope of $M$ verse $W$ is $0.877\pm0.085$ whereas the average slope from simulated data is $1.005\pm0.005$. For the first cut the average simulated slope was $-1.001\pm0.005$ which is in good agreement with the observations. Note that the only parameters that have been crudely estimated is the standard deviation used in the simulation.

\section{Selection of supernovae}
The above analysis qualitatively shows that the strong width verses redshift dependence shown in Figure \ref{f6} is not due to the expansion of the universe but it caused by magnitude and other selection effects that are due to the use of a distance modulus based on concordance  cosmology.  This section examines a simple selection model that provides a quantitative estimate of the selection bias.

Although the width, $W$, has sufficient range for selection to occur this cannot  explain Figure \ref{f6}. If there were no redshift dependence of the selection effects the distribution of observed widths at any redshift should cover the complete distribution of widths. In this case Figure \ref{f6} should show a scatter plot without any systematic variation with redshift. Clearly, there is a systematic variation of observed widths with redshift that needs to be explained.

What selection effects could produce the apparent dependence of width on redshift?  The technique for the supernova observations is a two-stage process \citep{Perlmutter03, Strolger04, Riess04}. The first stage is to conduct repeated observations of many target galaxies to look for the occurrence of supernovae. Having found a possible candidate the second stage is to conduct extensive observations of magnitude and spectra to identify the type of supernova and to measure its light curve. This second stage is extremely expensive of resources and it is essential to be able to determine quickly the type of the supernova so that the maximum yield of type 1a supernovae is achieved. Since current investigators assume that the type 1a supernovae have essentially a fixed absolute magnitude (with possible corrections for the stretch factor), one of the criteria they used is to reject any candidate whose predicted absolute magnitude (at maximum light) is outside a rather narrow range. The essential point is that the absolute magnitudes are calculated using concordance cosmology and hence the selection of candidates is dependent on the luminosity-distance modulus given by concordance cosmology.

An important quantity is the difference in the distance moduli, namely
\begin{equation}
\label{dm}
f(z)=\mu_{BB}(z)-\mu_{CC}(z),
\end{equation}
where $\mu_{CC}$ is the distance modulus for curvature-cosmology and $\mu_{BB}$ is that for concordance cosmology with $h=0.7$ and $\Omega _M=0.3$. For the range of redshifts used here the function $f(z)$ is approximately $1.37Z$, which is approximately equivalent to the removal of the redshift due to the time dilation from the distance modulus. Thus if, as we are assuming, the curvature cosmology distance modulus is correct the computed absolute magnitude using concordance cosmology will be larger than it should. Thus if a distant supernova is chosen to have the standard absolute magnitude using concordance cosmology then its true absolute magnitude will be weaker and its width will be wider.

In a comprehensive description of the selection procedure for a major survey \citet{Strolger04} state:
\emph{Best fits required consistency in the light curve shape and peak colour (to within magnitude limits) and in peak luminosity in that the derived absolute magnitude in the rest-frame B band had to be consistent with observed distribution of absolute B-band magnitudes shown in \citet{Richardson02}}.
If the supernovae have constant energy then if in addition there is any selection based on light curve width it can be treated as a selection based on magnitude.

In order to translate these selection criteria into a working model it is reasonable to assume that the probability that a supernova is selected will have a normal distribution of its magnitude. Considering that the Union project \citet{Kowalski08} is a amalgamation of several supernova projects the assumption of a normal distribution is even more reasonable.  Therefore if the magnitude has a normal distribution with a mean of $f(z)$ (Equation \ref{dm}) and a standard deviation of $\sigma$ and the selection process has a normal distribution with a mean of zero and a standard deviation of $\epsilon$ then the combined distribution for the probability of selection a supernova with absolute magnitude $M$ is
\begin{equation}
\label{dm1}
p(M)=\frac{1}{\sqrt{2\pi}\sigma\epsilon}\exp \left(-\frac{\sigma^2 + \epsilon^2}{2\sigma^2\epsilon^2}(M-M_0)^2-\frac{(f(z))^2}{\sigma^2 + \epsilon^2} \right ) ,
\end{equation}
where the mean magnitude is $M_0=\alpha f(z)$ with $\alpha=\sigma^2/(\sigma^2 + \epsilon^2)$ and its standard deviation is $\sigma\epsilon/\sqrt{(\sigma^2+\epsilon^2)}$. The simplest case and the one that best reflects the selection process is to assume that $\epsilon \propto \sigma$. The value of $\alpha$ was adjusted to provide a reasonable fit to the data shown in Figure \ref{f6}. The assumed value is $\alpha=0.79$ which gives $\epsilon=0.52\sigma$ and a standard deviation for $M$ of $0.46\sigma$.
The function $\alpha f(z)$ with $\alpha=0.79$ is shown as the dashed (red) curve in Figure \ref{f6}. Using the numbers in Table \ref{sna1} for all supernovae an  estimate of the standard deviation is $\alpha \sigma_M^2=(0.274)^2-(0.089)^2$ which gives $\sigma_M=0.29$ mag, which is less than $0.56$ mag  which \citet{Richardson02} quote for a much more heterogeneous mix of type 1a supernovae.  Thus the standard deviation of the selection distribution is $\epsilon=0.15$ mag. With this essentially constant energy model the light curve widths will have the same distribution which gives $\sigma_W=\sqrt{(0.153^2-0.066^2)/0.79}=0.16$. Since the magnitudes have additional variation due to uncertainties in extinction and distance the intrinsic standard deviation of the magnitudes is closer to 0.16 mag, which implies that the intrinsic standard deviation of the fluence is $\sqrt{2}\times 0.16=0.22$ mag. It is apparent that this simple model can provide a good fit to the data.

\section{Magnitudes of supernovae}
\label{s6.5.2}
\begin{figure}[!htb]
\includegraphics[width=0.99\textwidth,clip=true]{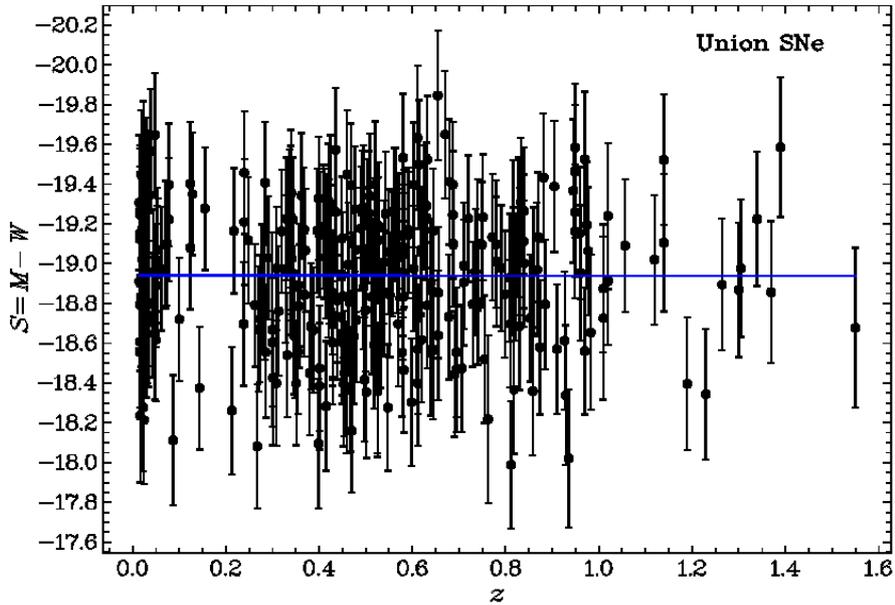}
\caption{Absolute fluences (i.e. absolute energies, $S_i=M_i-W_i$) verses redshift for Union data. Solid line (blue) is the line of best fit.\label{f9} }
\end{figure}

It has been argued that the fluence of type 1a supernovae makes an excellent `standard candle'.  Now we investigate whether the fluences of type 1a supernovae are independent of redshift.  Figure \ref{f9} shows the fluence, defined here for each supernova as  $S=M_{ref}=M-W$, of the Union supernovae as a function of redshift  $z$.

For a fit with $Z$ instead of $z$ the slope is $0.019\pm0.090$ which is in excellent agreement with zero. Thus showing that curvature cosmology can provide an excellent fit to the supernova data without the fitting of any free parameters. Overall the only assumption that is made that is specific to these supernovae is that $\epsilon \propto \sigma$ and that only effects the selection distribution.  Since there is no deviation from the straight line at large redshifts there is no need for dark energy \citep{Turner99} or quintessence \citep{Steinhardt98} both of which are meaningless in curvature-cosmology.

The estimate of the intercept, that is $M_{ref}$, is $-18.948\pm0.043$. \citet{Riess05} have measured accurate distances to two galaxies containing nearby supernovae. Together with two earlier measurements, they derive an absolute magnitude of type 1a supernovae of $-19.17\pm0.07$.  Hence the reduced Hubble constant, $h$, can be estimated from
\begin{equation}
-18.948 -5\log(0.7)=-19.17 -5\log(h)
\end{equation}
to get $h=0.632\pm0.024$. Thus using these values the measured Hubble constant is $63.2\pm2.4$ kms$^{-1}$ Mpc$^{-1}$. Note that the predicted value for the Hubble constant in curvature-cosmology is $H=64.4\pm 0.2\mbox{ km s}^{-1}\mbox{ Mpc}^{-1}$ which is in excellent agreement.

\section{Redshift distribution of supernovae}
This model predicts that the probability of detection of a supernova is given by integration of Equation (\ref{dm1}) over the magnitude to get
\begin{equation}
\label{dm2}
p_{sel}(M)=\frac{1}{\sqrt{2\pi}\eta}\exp(-\frac{(f(z))^2}{2\eta^2}),
\end{equation}
where as before it is assumed that $\epsilon=\sigma$ and therefore $\eta = \sigma /\sqrt{\alpha})$. In order to test this distribution the probability was computed for each supernova and the average probability was determine for each cut. The next step was to define a limiting magnitude which could be seen at nearly all the redshifts. The median apparent magnitude in the furthest cut was 24.72 mag. Thus all supernovae brighter than this should be seen in earlier cuts. The exceptions were two fainter supernovae in the second furthest cut. Restricting the analysis to supernovae brighter than 24.72 mag removes the first order dependence of selection on apparent magnitude.

Then the expected number of supernovae in each cut is the density multiplied by the volume and the average probability.  The value of $\sigma$ has to include all intrinsic and measurement variations in the observed magnitude. The computation was done with the estimate above of $\sigma=0.31$ mag. Thus, with $\alpha=0.7$ mag $\eta=0.37$ mag.  The results are shown for each cut in Table \ref{sna2}.

\begin{table}[!htb]
\begin{center}
\caption{Density distribution}\label{sna2}
\begin{tabular}{lcrccc}
\hline
No. & $z$& Vol. \tablenotemark{a}& Prob. \tablenotemark{b} & Exp. no. \tablenotemark{c} & Obs. \tablenotemark{d} \\
54 & 0.034 & 0.9   & 1.050 &  19.6 & $54\pm7$ \\
50 & 0.317 & 11.5  & 0.528 & 116.6 & $50\pm7$ \\
50 & 0.449 & 9.3   & 0.304 &  50.0 & $50\pm7$ \\
51 & 0.561 & 14.7  & 0.183 &  44.7 & $51\pm7$ \\
45 & 0.713 & 32.6  & 0.087 &  43.6 & $43\pm7$ \\
50 & 1.004 & 188.6 & 0.009 &  23.1 \tablenotemark{e} & $25\pm5$ \\
\hline
\end{tabular}
\end{center}
\tablenotetext{a}{Volume in Gpc$^{3}$ for $4\pi$ steradians}\\
\tablenotetext{b}{Average selection probability: Equation (\ref{dm2}) ($/eta =0.37$ mag)}\\
\tablenotetext{c}{Expected number: product of selection probability and density}\\
\tablenotetext{d}{The observed number with $m_B<24.72$ mag}\\
\tablenotetext{e}{This includes a 50\% factor to allow for the magnitude cutoff }\\
\end{table}

There is the implicit assumption in this calculation that the search time and search procedure are independent of redshift. The computed number was normalised so that the predicted number in the third cut was equal the observed number. Except for the first two cuts the agreement with the observed values is excellent. The poor agreement in the first two cuts is probably due to other observational and selection processes and has very little dependence  on the assumed cosmology.  However the good agreement for the outer cuts provides strong support for the model.  If this model is correct it predicts that it will be extremely difficult to find supernovae with redshifts greater than $z=1.5$ using the current concordance theory to select observations.

An identical analysis using concordance cosmology has similar volumes for each cut and assuming no selection bias it implies very strong evolution. For example with the computed number normalised using the third cut, the expected numbers (i.e. equivalent to column five in Table \ref{sna2}) are: 4, 58, 50, 82, 188, and 1108. Thus if concordance cosmology is correct the search time and the search procedures are roughly independent of redshift then the production of type 1a supernovae only occurs for redshifts less than about $z=2$.  Furthermore  the supernovae density has increased very rapidly since the start of production. Since normal galaxies and stars are seen at higher redshifts this shows an unusual evolution in supernovae production rates.

\section{Conclusions}
Assuming that curvature-cosmology is correct it has been shown that there is very strong support for the proposition that the most invariant property of type 1a supernovae is their total energy and not their peak magnitude. Given an essentially constant energy there is an inverse relationship between the peak luminosity and the width of the light curve. The total energy or fluence has an estimated standard deviation of 0.16 mag. Since the prime characteristic used for selecting these supernovae is the peak magnitude which is computed using concordance cosmology there is a strong bias that results in intrinsically weak supernovae being selected at higher redshifts. This is because the distance modulus in concordance cosmology is approximately $1.37Z$ larger than that for curvature-cosmology. Using a simple model for the selection process it was shown that it predicts the observed  dependence for the light curve widths on redshift. Furthermore an extension to the constant energy supernova model can explain the curious paradox that for nearby supernovae the magnitude-width slope has the opposite sign to the global slope. The conclusion is that with a simple selection model these supernovae observations are fully compatible with curvature-cosmology. This is strong support for the premise that there is no time dilation and hence no  universal expansion.

The strongest support for constant fluence as a function of redshift is shown by Figure \ref{f9}. It shows that without any fitted parameters the total energy or fluence of these supernovae is independent of redshift. Thus there is no experimental evidence for departures that have been ascribed to dark energy. The Hubble constant is computed to be $63.1\pm2.5$ km s$^{-1}$ Mpc$^{-1}$ which agrees with the curvature-cosmology prediction.

Finally these models in a static cosmology show good agreement with the observed density distribution of supernovae at high redshifts. Whereas the standard model with concordance cosmology requires unusual evolution of the supernovae production rates at redshifts greater than $z=0.5$.

\section{Acknowledgments}
This research has made use of the NASA/IPAC Extragalactic Database (NED) that is operated by the Jet Propulsion Laboratory, California Institute of Technology, under contract with the National Aeronautics and Space Administration. The graphics have been done using the DISLIN plotting library provided by the Max-Plank-Institute in Lindau. I thank the referee for constructive criticisms.

\section{Appendix: Curvature-cosmology}
Curvature-cosmology \citep{Crawford06,Crawford09a}  is a complete cosmology that shows excellent agreement with all major cosmological observations without needing dark matter or dark energy. It is compatible with both general relativity and quantum mechanics and obeys the perfect cosmological principle that the universe is statistically the same at all places and times. For a simple universal model of a uniform high temperature plasma it predicts that the plasma has a temperature of $2.56 \times 10^9 {\mbox{ K}}$. The theory has a good fit to the background X-ray radiation between the energies of 10-300 keV. The fitted temperature was $2.62\pm0.04\times 10^9$ K and the fitted density was equivalent to $1.55\pm0.01$ hydrogen atoms per cubic metre. For the simple homogeneous model this density is the only free parameter in the curvature-cosmology theory. Furthermore it predicts that the background microwave radiation should have a temperature of 3.18 K, and  that the Hubble constant (Equation \ref{e60}) is  $H=64.4\pm 0.2\mbox{ km s}^{-1}\mbox{ Mpc}^{-1}$. This new theory is based on two major hypotheses. The first is that the Hubble redshift  is due to an interaction of photons with curved spacetime. In this sense, it is a tired-light model. The important result is that the rate of energy loss (to extremely low energy secondary photons) is given by
\begin{equation}
\label{e1}
\frac{1}{E}\frac{dE}{ds} =-\left(\frac{8\pi GNM_H}{c^2} \right)^{1/2},
\end{equation}
where $M_H$ is the mass of a hydrogen atom and the effective density is $\rho=NM_H$. The Hubble constant is predicted to be
\begin{eqnarray}
\label{e60}
H &=&\frac{c}{E}\frac{{dE}}{{ds}} =-\left( 8\pi GM_H N \right)^{1/2} \nonumber \\
 & = & 1.671 \times 10^{ - 18} N^{1/2}\mbox{m}^{-1}\nonumber \\
 & = & 51.69 N^{1/2}\mbox{kms}^{-1}\mbox{ Mpc}^{-1}.
\end{eqnarray}
Thus the energy loss  is proportional to the integral of the square root of the density along the photon's path. The second hypothesis is that there is a pressure that acts to stabilise expansion and provides a static stable universe.

In this theory the distance travelled by a photon  is $R\chi$, where  $\chi =\ln(1+z)/\sqrt{3}$, and since the velocity of light is a universal constant the time taken is $R\chi/c$. There is a close analogy to motion on the surface of the earth with radius R. Light travels along great circles and $\chi$  is the angle subtended along the great circle between two points. The geometry of this curvature-cosmology is that of a three-dimensional surface of a four-dimensional hypersphere. It is identical to that for Einstein's static universe and is the same as that for the concordance model without expansion. For a static universe, there is no ambiguity in the definition of distances and times. One can use a universal cosmic time and define distances in light travel times or any other convenient measure. For this geometry the area of a sphere with radius $R$  is given by
\begin{equation}
\label{e2}
A(r) = 4\pi R^2 \sin ^2 (\chi ).
\end{equation}
The surface is finite and  $\chi$ can vary from 0 to $\pi$. Integration of this equation with respect to $\chi$  gives the volume $V$, namely,
\begin{equation}
\label{e3}
V(r) = 2\pi R^3 \left[ {\chi  - \frac{1}{2}\sin (2\chi )} \right].
\end{equation}
Clearly the maximum volume is $2\pi^2 R^3$ and  the theory predicts that
\begin{equation}
\label{e4}
R = \sqrt{\frac{{3c^2 }}
{{8\pi GM_H N}}}  = 3.100 \times 10^{26} N^{1/2} \mbox{ m}.
\end{equation}
Furthermore $R$ is related to the Hubble constant, $H$ by
\begin{equation}
\label{e5}
R = \frac{\sqrt{3} c}{H}.
\end{equation}
The distance modulus is obtained by combining the energy loss rate with the area equations to get
\begin{eqnarray}
\label{e8}
\mu_{CC}&=& m - M = 5\log \left[{\sqrt{3} \sin (\chi)} \right] + 2.5\log(1+z)\nonumber \\
&& - 5\log (h) + 42.384.
\end{eqnarray}
Since, by definition, the increase in bandwidth with redshift is included in  the K-correction there is only one power of $(1+z)$ included.


\begin{thebibliography}{999}
\bibitem[Astier et al.(2005)]{Astier05}Astier, A., Guy, J., Regnault, N., Pain, R., Aubourg, E., Balam, D., Basa, S., Carlberg, R. G., Fàbbro, S., Fouchez, D., Hook, I.M., Howell, D. A., Lafoux, H., Neill, J. D., Palanque-Delabrouille, N., Per?et, K., Pritchet, C.J., Rich, J., Sullivan, M., Taillet, R., Aldering, G., Antilogus, P., Arsenijevic, V., Balland, C., Baumont, S., Bronder, J., Courtois, H., Ellis, R. S., Filiol, M., Gonçalves, A. C., Goobar, A., Guide, D., Hardin, D., Lusset, V., Lidman, C., McMahon, R., Mouchet, M., Mourao, A., Perlmutter, S., Ripoche, P., Tao, C. \& Walton, N., 2005, \aap, 207, 1504
\bibitem[Crawford(2006)]{Crawford06} Crawford, D. F., 2006, Curvature Cosmology, Boca Ratan, BrownWalker Press
\bibitem[Crawford(2009a)]{Crawford09a} Crawford, D. F., 2009a, (Contains second edition of the book `Curvature Cosmology'), http://www.davidcrawford.bigpondhosting.com
\bibitem[Crawford(2009b)]{Crawford09b} Crawford, D. F., 2009b, arXiv:0901.4169
\bibitem[Davis et al. (2007)]{Davis07}Davis, T. M.; M\"{o}rtsell, E., Sollerman, J., Becker, A. C., Blondin, S., Challis, P., Clocchiatti, A., Filippenko, A. V., Foley, R. J., Garnavich, P. M., Jha, S., Krisciunas, K., Kirshner, R. P., Leibundgut, B., Li, W., Matheson, T., Miknaitis, G., Pignata, G., Rest, A., Riess, A. G., Schmidt, B. P., Smith, R. C.,                 Spyromilio, J., Stubbs, C. W., Suntzeff, N. B., Tonry, J. L., Wood-Vasey, W. M. \& Zenteno, A., 2007, \apj, 666, 716
\bibitem[Goldhaber et al.(2001)]{Goldhaber01}Goldhaber, G., Groom, D. E., Kim, A., Algering, G., Astier, P., Couley, A., Deustua, S. E., Ellis, R., Fabbro, S., Fruchter, A. S., Goobar, A., Hook, I., Orwin, M., Kim, M., Knop, R. A., Lidman, C., McMahon, R., Nugent, P. E., Pain, R., Panagia, N., Pennypacker, C. R., Perlmutter, S., Ruiz-Lapuente, P., Schaefer, B., Walton, N. A. \& York, T., 2001, \apj, 558, 359
\bibitem[Guy et al.(2005)]{Guy05}Guy, J., Astier, P., Nobili, S., Regnault, N. \& Pain, R., 2005, \aap, 443, 781
\bibitem[Hamuy et al.(1996)]{Hamuy96}Hamuy, M., Phillips, M. M., Schommer, R. A. \& Suntzeff, N. B., 1996, \aj, 112, 2391
\bibitem[Hillebrandt \& Niemer(2000)]{Hillebrandt00}Hillebrandt, W. \& Niemer, J. C.,  2000, \araa, 38, 191
\bibitem[Knop et al.(2003)]{Knop03}Knop, R. A., Aldering, G., Amanullah, R., Astier, P., Blanc, G., Burns, M. S., Conley, A., Deustua, S. E., Doi, M., Ellis, R., Fabbro, S., Folatelli, G., Fruchter, A. S., Garavini, G., Garmond, S., Garton, K., Gibbons, R., Goldhaber, G., Goobar, A., Groom, D.E., Hardin, D., Hook, I., Howell, D.A., Kim, A. G., Lee, B. C., Lidman, C., Mendez, J., Nobili, S., Nugent, P. E., Pain, R., Panagia, N., Pennypacker, C. R., Perlmutter, S., Quimby, R., Raux, J., Regnault, N., Ruiz-Lapuente, P., Sainton, G., Schaefer, B., Schahmabeche, K., Smith, E., Spadafora, A. L., Stanishev, V., Sullivan, M., Walton, N. A., Wang, L., Wood-Vasey, W. M. \& Yasuda, N., 2003, \apj, 598, 102
\bibitem[Kowalski et al.(2008)]{Kowalski08}Kowalski, M., Rubin,D., Aldering,G., Agostinho, R. J., Amadon, A., Amanullah, R., Balland, C., Barbary, K., Blanc, G., Challis, P. J., Conley, A., Connolly, N. V., Covarrubias, R., Dawson, K. S., Deustua, S. E., Ellis, R., Fabbro, S., Fadeyev, V., Fan, X., Farris, B., Folatelli, G., Frye, B. L., Garavini, G., Gates, E. L., Germany, L., Goldhaber, G., Goldman, B., Goobar, A., Groom, D. E., Haissinski, J., Hardin, D., Hook, I., Kent, S., Kim, A. G., Knop, R. A., Lidman, C., Linder, E. V., Mendez, J., Meyers, J., Miller, G. J., Moniez, M., Mourao, A. M., Newberg, H., Nobili, S., Nugent, P. E., Pain, R., Perdereau, O., Perlmutter, S., Phillips, M. M., Prasad, V., Quimby, R., Regnault, N., Rich, J., Rubenstein, E. P., Ruiz-Lapuente, P., Santos, F. D., Schaefer, B. E., Schommer, R. A., Smith, R. C., Soderberg, A. M., Spadafora, A. L., Strolger, L. -G., Strovink, M., Suntzeff, N. B., Suzuki, N., Thomas, R. C., Walton, N. A., Wang, L., Wood-Vasey, W. M. \& Yun, J. L., 2008, /apj, 686, 749
\bibitem[Perlmutter \& Schmidt(2003)]{Perlmutter03}Perlmutter, S. \& Schmidt, B. P. (2003) in Supernovae, Gamma Ray Bursts, K. Weiler, ed. (Springer: New York)
\bibitem[Perlmutter et al.(1999)]{Perlmutter99}Perlmutter, S., Aldering, G., Goldhaber, G., Knop, R.A., Nugent, P., Castro, P.G., Deustua, S., Fabbro, S., Goobar A., Groom, D. E., Hook, I. M., Kim, A. G., Kim, M, Y., Lee, J. C, Nunes, N. J., Pain, R., Pennypacker, C. R., Quimby, R., Lidman, C., Ellis, R. S., Irwin, M., McMahon, R. G., Ruiz-Lapuente, P., Walton, N., Shaefer, B., Boyle, B. J., Filippenko, A. V., Matheson, T., Fruchter, A. S., Panagia, N., Newberg, H. J. M. \& Couch, W. J., 1999, \apj, 517, 565
\bibitem[Phillips(1993)]{Phillips93}Phillips, M. M., 1993, \apj, 413, L105
\bibitem[Press et al.(1992)]{Press92}Press, W. H., Teukolsky, A. A., Vetterling, W. T. \& Flannery, B. P., 1992, Numerical Recipes in Fortran 77, Cambridge University Press
\bibitem[Richardson et al.(2002)]{Richardson02}Richardson, D., Branch, D., Casebeer, D., Millard, J., Thomas, R. C. \& Baron, E., 2002, Astron. J. 123, 745
\bibitem[Riess et al.(2004)]{Riess04}Riess, A. G., Strolger, L. -G., Tonry, J., Casertano, S., Fergusen, H. C., Mobasher, B., Challis, P., Filippenko, A. V., Jha, S., Li, W., Chornock, R., Kirshner, R. P., Leibundgut, B., Dickinson, M., Livio, M., Giavalisco, M., Steidel, C. C., Benitez, N. \& Tsvetanov, Z., 2004, \apj, 607, 645
\bibitem[Riess et al.(2005)]{Riess05}Riess, A. G., Li, W., Stetson, P. B., Filippenko, A. V., Jha, S., Kirshner, R. P., Challis, P. M., Garnavitch, P. M. \& Chornock, R., 2005, \apj, 627, 579
\bibitem[Steinhardt  \& Caldwell(1998)]{Steinhardt98}Steinhardt, P. J. \& Caldwell, R. R.,  1998, Cosmic Microwave Background \& Large Scale Structure of the Universe, Astr. Soc. Pacif. Conf. Ser. 151, 13
\bibitem[Strolger et al.(2004)]{Strolger04}Strolger, L. -G., Riess, A. G., Dahlen, T., Livio, M., Panagia, N., Challis, P., Tonry, J. L., Filippenko, A. V., Chornock, R., Fergusen, H., Koekemoer, A., Mobasher, B., Dickinson, M., Giavalisco, M., Casertano, S., Hook, R., Blondin, S., Leibundgut, B., Nonino, M., Rosati, P., Spinrad, H., Steidel, C.C., Stern, D., Garnavich, P. M., Matheson, T., Grogin, N., Hornschemeir, A., Kretchmer, C., Laidler, V. G., Lee, K., Lucas, R., de Mello, D., Moustakas, L. A., Ravindranath, S., Richardson, M. \& Taylor, E., 2004, \apj, 613, 200
\bibitem[Turner(1999)]{Turner99}Turner, M. S., 1999. in The Third Stromlo Symposium: The Galactic Halo, Gibson, B. K., Axelrod, T. S. \& Putman, M. E. (eds), Astr. Soc. Pacific Conf. Ser. 165, 431
\bibitem[Wood-Vasey et al. (2007)]{Wood-Vasey07}Wood-Vasey, W. M.,; Friedman, A. S., Bloom, J. S., Hicken, M., Modjaz, M., Kirshner, R. P., Starr, D. L., Blake, C. H., Falco, E. E., Szentgyorgyi, A. H., Challis, P., Blondin, S. \& Rest, A., 2008, 689, 377.
\end{thebibliography}
\end{document}